\documentclass[12pt,showpacs,preprintnumbers]{revtex4}
\usepackage{graphicx}% Include figure files
\usepackage{dcolumn}% Align table columns on decimal point
\usepackage{bm}% bold math
\usepackage{amssymb}
\usepackage{epsfig}    
\usepackage{here}
\def\beq{\begin{equation}}
\def\eeq{\end{equation}}
\def\ba{\begin{array}}
\def\ea{\end{array}}
\def\bea{\begin{eqnarray}}
\def\eea{\end{eqnarray}}

\def\sq2{\sqrt{2}}
\def\End{\end{document}}
\begin{document}                                                              
%\draft

\title{Higgs FCNC $h\to t^* c$ decay at the Muon collider}%
\author{%
{G.~Tetlalmatzi\,$^1$},~~{J.G.~Contreras\,$^2$},~~{F.~Larios\,$^2$},
~~{M.A.~P\'erez\,$^3$}  
}
\affiliation{%
%\address{\vspace*{5mm}
\vspace*{2mm} 
$^1$Departamento de Ciencias B\'asicas, 
UPIBI-IPN, C.P. 07340, M\'exico D.F., M\'exico\\
$^2$Departamento de F\'{\i}sica Aplicada, 
CINVESTAV-M\'erida, A.P. 73, 97310 M\'erida, Yucat\'an, M\'exico\\
$^3$Departamento de F\'{\i}sica, CINVESTAV, A.P. 14-740, 07000,
M\'exico D.F., M\'exico
}
%\maketitle

%\vspace*{5mm} 
\begin{abstract}
\hspace*{-0.35cm}

We study the discovery potential of the flavor-changing neutral
coupling (FCNC) $htc$ of the Higgs boson and the top quark through
the rare tree-body decay $h \to Wbc$ at Muon colliders for a light
Higgs boson with mass $114 \leq m_h \leq 145$ GeV. This
decay mode may compete with the SM background induced by
the $hWW$ coupling in some models with a tree-level $htc$ coupling
and with models that predict this coupling at the one-loop level
in the range $10^{-2} \;-\; 10^{-1}$.  A future muon
collider could test the scalar FCNC decay $t\to hc$ via
Higgs decay $h\to t^* c \to b W^+ c$ down to values of the
coupling $g_{tc}=0.5$ (that are equivalent to 
BR$(t\to hc)\sim 5 \times 10^{-3}$).  The LHC could probe
values of $g_{tc}$ one order of magnitude smaller, unless
other processes beyond the SM appear that through intense
multi jet activity may clutter the $t\to hc$ signal.
%\hfill   ~~ [ Jul, 2009 ] }

\end{abstract}
%\vskip 1pc]

\maketitle

\setcounter{footnote}{0}
\renewcommand{\thefootnote}{\arabic{footnote}}

\section{Introduction}

Top quark decays induced by flavor-changing neutral couplings (FCNC) 
constitute a direct and sensitive probe of new-physics effects at
energy scales of a few hundred GeV \cite{topreviews,jaumerev}.
While this type of process is highly suppressed in the Standart Model
(SM) due to the GIM mechanism \cite{lorenzo,eilam}, most of its
extensions predict branching ratios for these decay modes that may
be observable at the LHC \cite{cordero}. 
It has been suggested also that some Higgs-mediated FCNC top quark
processes could be observed at the LHC \cite{branco,koji} and linear
colliders \cite{sher}. These processes are considered to be the
most likely to shed light on the nature of the electroweak symmetry
breaking mechanism if the FCNC top quark-Higgs boson couplings are
as big as expected in most models \cite{jaumerev,cordero}.
Our general aim in this paper is to study the possibility of
detecting the $htc$ FCNC interaction at Muon colliders. There 
are studies focused on FCNC Higgs decays in this kind of colliders
via its heavy lepton channels \cite{lorenzo2}.
The CERN LHC is expected to discover a light Higgs boson with a
mass $114 \leq m_h \leq 145$ GeV \cite{flaecher} and will determine
some of its properties such as its fermionic and bosonic decay modes
and couplings \cite{belyaev}. 
A $e^+e^-$ linear collider with a center of mass energy above
500 GeV will be able to significantly improve
these preliminary measurements \cite{aguilar}.
In particular, it has been suggested that the decay $t\to hc$ may be
detected if the Higgs boson is somewhat lighter than the top quark in
a linear collider with 500 fb$^{-1}$ integrated luminosity and
$\sqrt{s}=500$ GeV for models that predict branching ratios of order 
$10^{-4}$ for this FCNC decay mode \cite{sher}. 
In the present paper we address the question concerning the
possibility that the FCNC $htc$ coupling could be measured with
a higher accuracy in a muon collider through the three-body decay
mode $h\to t^*c\to Wbc$ for a light Higgs boson. In the SM, this
decay mode is determined by the $W$ pair production channel 
$h\to WW^* \to Wbc$ but we expect that the $htc$ coupling is high
enough to compete with the SM contribution.

It has been estimated that a linear collider with 500 fb$^{-1}$
integrated luminosity and a center of mass energy of 500 GeV
will begin to be sensitive to the $htc$ coupling through the
decay $t\to hc$ for extensions of the SM that predict branching
ratios of this decay mode of order $10^{-5} \;-\; 5\times 10^{-4}$.
This correspond to a range of the $htc$ coupling constant 
$\lambda_{tc}$ of order $0.06 \;-\; 0.4$ 
($g_{tc}$ between $0.02$ and $0.1$) \cite{sher}.
Models with a tree-level $htc$ coupling have branching ratios of
the order of $10^{-3} \;-\; 10^{-2}$, while models that induce this
coupling at the one loop level predict branching ratios of order
$10^{-5} \;-\; 10^{-4}$ for this FCNC channel
\cite{topreviews,jaumerev}.  In particular, in the context of the
MSSM the scalar $t\to hc$ decay would be greater than other
channels like $t\to cg$, and SUSY-QCD effects could yield
$BR(t\to hc) \sim 5\times 10^{-4}$\cite{jaumesusy}.
Similar values are possible in the context of the two-Higgs-doublet
model\cite{jaumethdm}.  
At the LHC, discovery limits for this decay mode were calculated in
Ref.~\cite{branco} and are of the order $5\times 10^{-5}$,
but the hadronic background may complicate the analysis.
In the case of the muon collider, we expect that the production
cross section on resonance for the chain
$\mu^+ \mu^- \to h \to t^* c \to Wbc$ may be high enough for some
extensions of the SM in such a way that it can compete with the SM
background induced by the $hWW^*$ coupling. We assume that a light
Higgs boson will be observed at the LHC and examine in detail the
discovery potential for its FCNC $htc$ coupling at a Muon collider.

At the LHC, unfortunately it is quite difficult to  observe the
three-body decay $h \to Wbc$ due to the fact that the expected QCD
background coming from 4 jet events is orders of magnitude larger
than the signal coming from the $htc$ or $hWW$ vertices.
In this case it will be necessary to
look for the purely leptonic events in order to have a discovery
channel for a light Higgs boson at the LHC \cite{gonzalez}.

%%%%%%%%%%%%%%%%%%%%%%%%%%%%%%%%%%%%%%%%%%%%%%%%%%%%%%%%%%%%%%%%%%%%%%%

\section{S-channel Higgs production at the Muon Collider}

A Muon collider designed with the purpose of studying the
Higgs resonance with a very fined tuned center-of-mass (CM)
energy equal to $m_h$
could become an effective Higgs factory where precise measurements
of the Higgs mass and width could be achieved\cite{taohan}.
It is expected that such a machine could yield annual integrated
luminosities of order $20 fb^{-1}$.  However, the luminosity depends
on how broad the muon energy spectrum is allowed to be.
Here we assume a very low (albeit supposedly feasible) energy spread
of a few MeVs and an integrated luminosity of order $1 fb^{-1}$.

Given the very narrow width of a light (below the $WW$ threshold)
Higgs boson the production cross section is extremely sensitive
to the energy of the colliding muons\cite{taohan}.
The Muon collider is expected to have a Gaussian shape of the
energy spectrum of the colliding beams with an rms deviation
\bea
R = \frac{\sqrt{2}\sigma_E }{\sqrt{s}} \, .
\eea 
Where $\sigma_E$ is the energy spread of the beams, and
$\sqrt{s}$ is their CM energy. Notice that for $\sqrt{s}=10^2$
GeV and $\sigma_E = 7$ MeV the rms deviation is $R=10^{-4}$.
Near the resonance region
the production cross section of the s-channel process
$\mu^- \mu^+ \to h \to X$ is
\bea
\sigma (\sqrt{s}) = \frac{4\pi \Gamma (h\to\mu^- \mu^+) 
\Gamma (h\to X)}{(s-m^2_h)^2 + m^2_h \Gamma_h^2}
\eea
where $\Gamma_h$ is the total width of the Higgs boson and
$s=(p_{\mu^-} +p_{\mu^+})^2$ is the CM energy squared of the
colliding muons.  Neglecting Bremmstrahlung, we can write the
effective cross section as the convolution of the cross section
above with the Gaussian distribution of the CM energy centered
at $\sqrt{s} = m_h$\cite{taohan}:
\bea
\bar \sigma = \frac{1}{\sqrt{2\pi}\sigma_E} \, \int \sigma (x) 
exp \left( {- \frac{(x-m_h)^2}{2\sigma^2_E}}\right) \, dx
\eea

We can re-write the above formula as:
\bea
\bar \sigma = \frac{4\pi}{\sqrt{\pi}} 
BF(h\to \mu \mu) BF(h\to X) \; 
\frac{a^2}{Rm^2_h} \int^{tmax}_{tmin} 
\frac{e^{-t^2/R^2}}{a^2+t^2 (t+2)^2} \; dt
\label{sigmahtox}
\eea
where $a\equiv \Gamma_h/m_h$ and the limits of integration
$tmax (tmin)$ have been taken as $\pm 10^{-2}$.  These limits
cover all the significant contribution from the Gaussian
distribution when we consider values of $R=10^{-4}$ and
$a\sim 10^{-4}$ for the rms deviation and the width-to-mass
ratio of the light Higgs respectively.

\begin{figure}
\includegraphics[width=9.0cm,height=7.5cm]{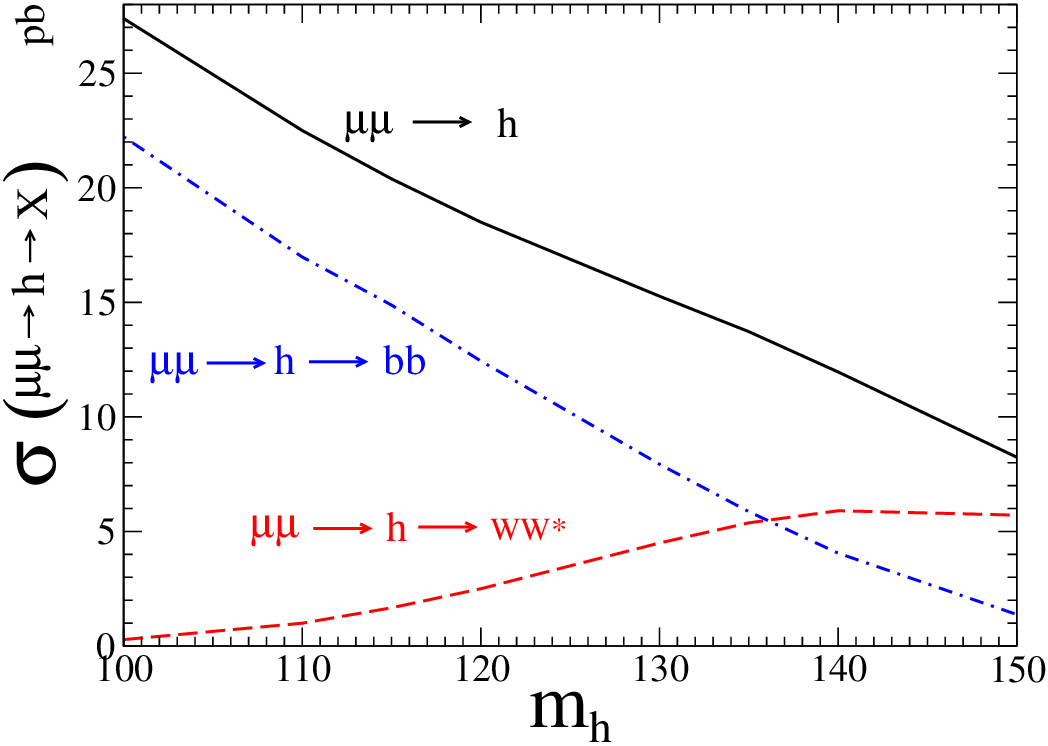}
%\vspace*{-2mm}
\caption{The total $\mu^+\mu^- \to h$, $\mu^+\mu^- \to b\bar b$
and $\mu^+\mu^- \to h\to W^+ {W}^{*-}$ cross section for a Muon
collider operating at $\sqrt{s} = m_h$ with an energy
spread $R=10^{-4}$ ($\Delta E \sim 7$MeV).}
\label{sigmatotrm4}
\end{figure}

In Fig.~\ref{sigmatotrm4} we show the Muon collider resonant Higgs
production:  the two most important channels 
$\mu^+ \mu^- \to h \to b\bar b$ and
$\mu^+ \mu^- \to h \to W^+ W^{*-}$ as well as the total, which is
obtained by setting $BF(h\to X) \equiv 1$ in Eq.~(\ref{sigmahtox}).
As the Higgs mass approaches the 
Since the $BF( W^- \to b\bar c )$ is of order
$0.7 \times |V_{cb}|^2 = 0.7 \times 0.04^2 \sim 10^{-3}$ we
expect the  $\mu^+ \mu^- \to h \to W^+ b\bar c$ channel to be
of order a few fb.

\begin{figure}
\includegraphics[width=9.0cm,height=7.5cm]{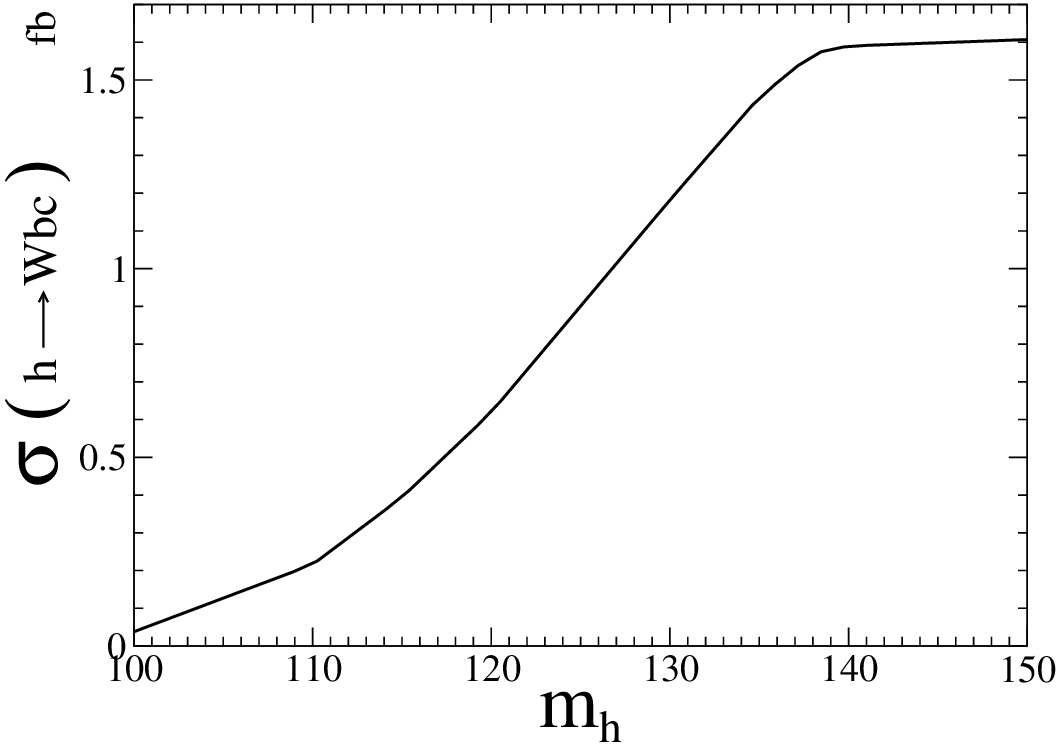}
\vspace*{2mm}
\caption{The effective $\sigma (h\to W^+b\bar c)$ cross section
for a Muon collider operating at $\sqrt{s} = m_h$ with a energy
spread $R=10^{-4}$ ($\Delta E \sim 7$MeV).}
\label{smhwbc}
\end{figure}

In Fig.~\ref{smhwbc} we show the resonant Higgs production
cross section of a light Higgs decaying to the $W^+b\bar c$
state.  An additional reduction factor of order 0.6 could be
used to account for Bremmstrahlung\cite{taohan}.
Fig.~\ref{smhwbc} shows a very small cross section of order
$1 fb$ that would hardly yield one event with the integrated
luminosity we consider.  Comparing with the more general
$h\to Wjj$ decay channel the $Wbc$ contains a CKM $V_{cb}$
factor of order $10^{-3}$ that explains why this SM decay of
the light Higgs is so suppressed.

In other words, the $Wjj$ mode is $10^3$ times higher and can
yield a significant number of events. In fact, the $h\to WW^*$
mode turns out to be as good probe as $h\to bb$ even for a
light Higgs (the latter is a dominant decay channel but it has
a substantial background)\cite{taohan}.

The effective $htc$ coupling we use is defined as:
\bea
{\cal L} = \frac{g}{2\sqrt{2}} g_{tc} h\bar t c \; +\; h.c. 
\label{htc}
\eea
The effective coupling $g_{tc}$ is very small in the SM
$g^{SM}_{tc} \sim 10^{-6}$ but could be several orders of
magnitude higher in other models like the THDM where
an ansatz $g^{SM}_{tc} \sim \sqrt{m_t m_c}/M_W \sim 0.2$
is considered assuming the FC couplings scale with the quark
masses\cite{Cheng}.  It has recently been estimated that the LHC could
measure $g_{tc}$ down to the order of $0.04$\cite{branco}.

\begin{figure}
\includegraphics[width=9.0cm,height=7.5cm]{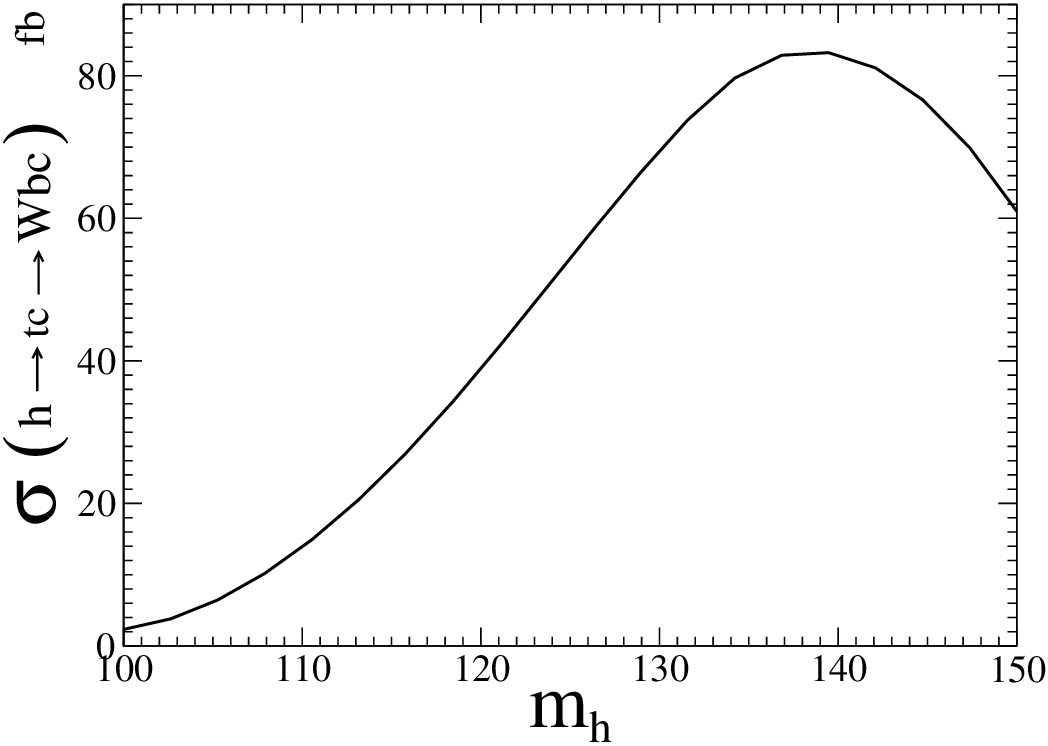}
\vspace*{2mm}
\caption{The effective $\sigma (h\to tc\to W^+b\bar c)$ cross
section for a Muon collider operating at $\sqrt{s} = m_h$ with
an energy spread $R=10^{-4}$ ($\Delta E \sim 7$MeV).}
\label{sigmatcrm4}
\end{figure}

In Fig.~\ref{sigmatcrm4} we show the production cross section
for the FC process $\mu^+ \mu^- \to h \to t \bar c \to W^+ b\bar c$
for a range of the Higgs mass.  There, the coupling $g_{tc}$ is
taken as $g_{tc}=1$.  For this size the contribution from
the SM $\mu^+ \mu^- \to h \to WW^* c \to W^+ b\bar c$ is about
two orders of magnitude smaller.  Fig.~\ref{sigmatcrm4} does
not include the SM process; if we take $g_{tc}$ of order $0.1$
the two contributions become similar and we would have to
include interference effects.  However, as cross sections of
order a few $fb$ are deemed to provide too little statistics
we can realize the coupling $g_{tc}$ has to be a least of
order $0.5$ to yield enough production.  In this respect,
the LHC has a much better sensitivity to this coupling than
the Muon collider.

\section{A Muon collider vs the LHC}

As mentioned before, for a Higgs mass smaller than the Top mass,
the decay $t\to hc\to b\bar b c$ could be used to measure the
coupling $g_{tc}$.    With the high statistics of top pair events
at the LHC a very good sensitivity could be reached for this
coupling; Ref.~\cite{branco} obtained a potential limit
$g_{tc}\leq 0.04$ for a total integrated luminosity of 100 $fb^{-1}$.

However, this limit could be greatly weakened if other (beyond the SM)
processes came into play that would produce the same experimental
signature as the one studied by Ref.~\cite{branco}.  Such signature
is based on a final state $l\nu bb\bar b j$ where a $b\bar b$ pair
comes from the decay of the Higgs, and the other $b$ jet (as well as
the lepton and neutrino) comes from the decay of the $\bar t$ quark.

Let us assume that supersymmetric partners are discovered at the LHC.
It is well known that multi-jet events with high missing transverse
energy are among the typical signals of these new resonances.  
Such is the case for some mSUGRA scenarios like the test points
LM1 or LM3 used by the CMS Collaboration to evaluate their
physics performance (see Chapter 13 in \cite{cms}).
Some of these processes have a total cross section around 50 pb
that is one order of magnitude smaller than the total cross section
for $t\bar t$ production.
The analysis of Ref.~\cite{branco} requires high $P_T$ for the jets,
leptons and missing $P_T$, and by doing this they cut out a lot of
the $t\bar t$ cross section (as well as other processes with lower
transverse momenta).  On the other hand, such high $P_T$ cuts are not
expected to affect as much mSUGRA processes which have very high
transverse momenta. A detailed quantitative study of this question is
beyond the scope of this paper. However, our claim is that
the presence of this kind of background would not invalidate their
analysis, but  it would potentially weaken their limits.

%%%%%%%%%%%%%%%%%%%%%regresar%%%%%%%%%%%%%%%%%%%%%%%%%%%%%%%%%%%%%

\section{Conclusions}

The Higgs FCNC $h\to t^* \bar c \to W^+ b \bar c$
may compete with the SM background induced by
the $hWW$ coupling in some models with a tree-level $htc$ coupling
and with models that predict this coupling at the one-loop level
in the range $10^{-2} \;-\; 10^{-1}$.
For a light Higgs boson of mass around 120 GeV a future muon
collider could test the scalar FCNC decay $t\to hc$ via
Higgs decay $h\to t^* c \to b W^+ c$ down to values of the
coupling $g_{tc}=0.5$ (that are equivalent to 
BR$(t\to hc)\sim 5 \times 10^{-3}$).  The LHC could probe
values of $g_{tc}$ one order of magnitude smaller, unless
other processes beyond the SM appear that through intense
multi jet activity may clutter the $t\to hc$ signal.

%%%%%%%%%%%%%%%%%%%%%%%%%%%%%%%%%%%%%%%%%%%%%%%%%%%%%%%%%%

\vspace*{4cm}

\noindent
{\bf Acknowledgements}~~~
We thank J.L. Diaz Cruz and R. Martinez for useful discussions. 
We acknowledge support from Conacyt (Mexico).

%%%%%%%%%%%%%%%%%%%%%%%%%%%%%%%%%%%%%%%%%%%%%%%%%%%%%%

%\newpage

%\vspace*{-2mm}

\end{document}